\documentclass{article}
\usepackage{spconf,amsmath,graphicx,hyperref,amssymb,enumitem,booktabs,minted,xcolor}

\usepackage[symbol]{footmisc}

\newcommand{\blfootnote}[1]{%
  \begingroup
  \renewcommand\thefootnote{}\footnote{#1}%
  \addtocounter{footnote}{-1}%
  \endgroup
}

\definecolor{lightmode-bg}{HTML}{fafaf9} % light mode bg
\setminted{
    fontsize=\footnotesize,
    breaklines=true,
    frame=lines,
    framesep=2mm,
    baselinestretch=1.2,
    bgcolor=lightmode-bg, 
}

\title{A Study on Improving Multi-class Audio Source Separation Via Decoupled CLAP Query Optimization and an Automated Data Engine}
\name{
Amirhossein Hajavi, 
Hanhee Lee $^*$,
Pushya Jain $^*$, 
Sky Qiao,
Emmanuel Ko, 
Yuanhao Yu,
Irina Kezele
}
\address{
Noah's Ark Lab, Huawei Technologies Canada co. \\
{\{amirhossein.hajavi, irina.kezele\}@huawei.com}
}
\begin{document}
%\ninept
%
\maketitle
\begin{abstract}
Language-queried audio source separation (LASS) enables extracting any sound source using natural language. However, adapting LASS models to application-specific sound classes is challenging due to noisy training data and limited semantic coverage of the CLAP-based control signals. We propose a framework comprised of an automated data engine for training-data curation and a two-stage optimization process for class-specific CLAP control signals. Our objective evaluations across seven sound classes show that data refinement and control signal optimization consistently improve source separation performance. Subjective evaluation with 17 participants further demonstrates perceptual improvements of the model trained with optimized control signals over baseline and similar commercial models.

\end{abstract}
\blfootnote{$^*$ Equal contribution}

\section{Introduction}
\label{sec:intro}

Universal audio source separation aims to isolate a mixture of multiple arbitrary sound sources to its components \cite{kavalerov2019universal}. Among its various paradigms, language-queried audio source separation (LASS) uses natural language as the control signal to extract a target sound source. The emergence of Contrastive Language-Audio Pre-training (CLAP) \cite{elizalde2023clap} has led to considerable advancements in the performance of LASS models \cite{ma2024clapsep, mahmud2024weakly, chung2024language, ma2025language, feng2025hybrid, yang2026clapmamba, liu2024separate, saijo2025leveraging, zhang2026mars}. By providing semantically aligned language and audio embeddings, CLAP enables effective language-based control signals, resulting in consistent improvements in source separation performance across both objective and subjective evaluations.

Audio source separation models are often deployed in application domains where a subset of sound classes is of primary interest. As a consequence, improving the separation quality of these specific classes becomes particularly important. Historically, application-specific source separation has primarily focused on tasks such as music source separation \cite{limited_defossez2019music, limited_rouard2023hybrid, limited_luo2023music, limited_kim2021kuielab} or target speech separation \cite{limited_luo2019conv}. The flexibility afforded by text encoders such as CLAP enables arbitrary sound classes to be specified using natural language rather than predefined label sets. Consequently, pretrained LASS models can be adapted to a much broader range of downstream applications \cite{limited_arbitrary_chen2024mdx}.

Identifying suitable target sound classes for downstream applications remains challenging. Although LASS models perform inference using natural language prompts, their performance on a target class depends heavily on how well that class is represented in the training data. In some cases where a general set of sound classes are selected as target sound classes, the collective samples of audio that are subcategories of the general classes can be used for training the source separation model. However, embeddings learned solely through CLAP do not necessarily preserve the hierarchical relationships between sound classes therefore can not be used effectively as control signals. 
% On the other hand, deploying large-scale CLAP models on edge devices brings forth a trade-off between the performance of the source separation and the latency of the entire pipeline. 

Existing work primarily improves source separation performance through architectural modifications to model training and inference \cite{limited_arbitrary_chen2024mdx, zhang2026domain}. Meanwhile, the challenges of curating useful training data and optimizing language-based control signals remain underexplored. Motivated by these limitations, we make the following contributions: (1) We propose a multi-stage automated data engine for collecting useful training data from the available large-scale datasets and validating their relevance to the target classes. (2) We propose an enhanced control signal by aggregating multiple CLAP embeddings for each class and decoupling the optimization of the resulting embeddings from the end-to-end pipeline to increase inter-class discriminability for general sound source categories. (3) Through extensive experiments on the 7 sound classes, we show that both our data engine and control-signal optimization significantly improve the performance of the source separation models for the target classes.

\begin{figure*}[!ht]
    \centering
    \includegraphics[width=0.80\linewidth]{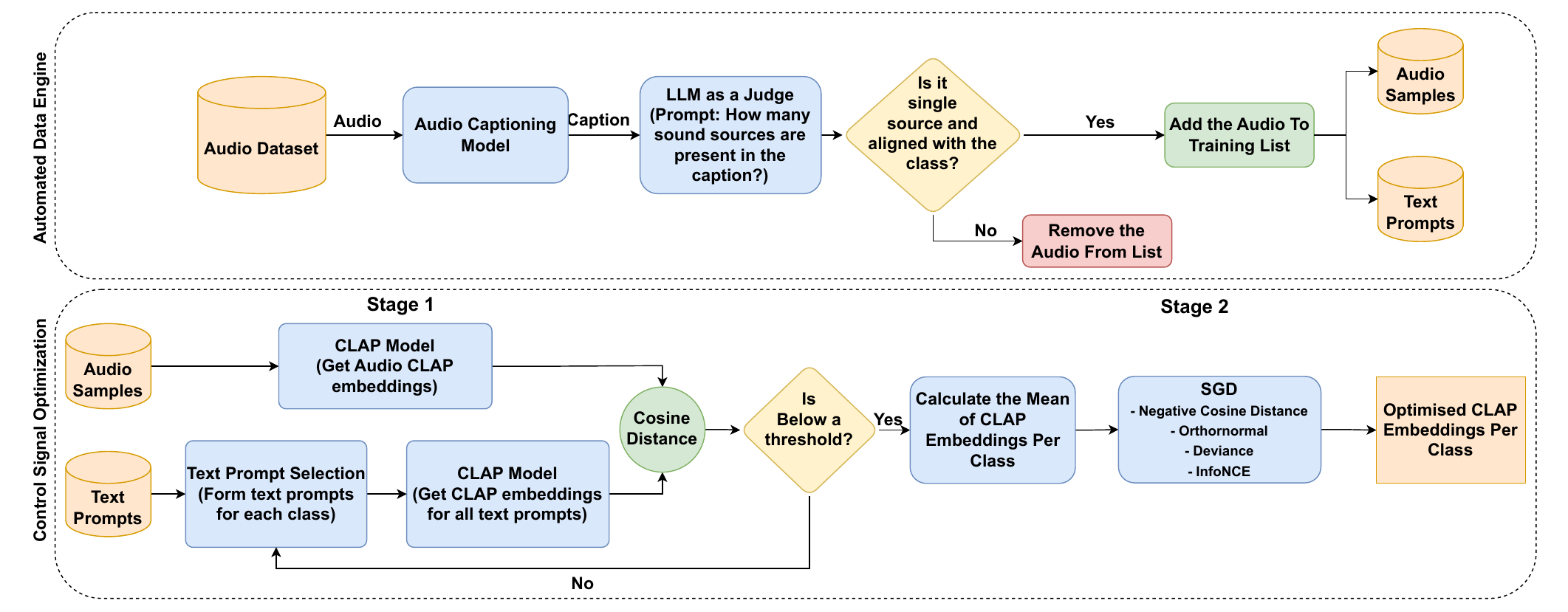}
    \caption{\footnotesize An overview of the proposed framework. (Top) The pipeline for the automated data engine which consists of an audio captioning and LLM-as-a-Judge component. (Bottom) The optimization process for control signals using an iterative procedure to select the candidate CLAP embeddings followed by a SGD optimizer.}
    \label{fig:overview}
\end{figure*}

\section{Related Work}

Early works on sound source separation focused on music source separation \cite{limited_defossez2019music, limited_rouard2023hybrid, limited_luo2023music, limited_kim2021kuielab} or target speech separation \cite{limited_luo2019conv}. The benchmarks defined for these tasks either included a predefined set of sound classes (e.g. four musical instrument classes in \cite{musdb18-hq}) or a reference speech signal from the target speaker \cite{cosentino2020librimix}. Universal audio source separation, on the other hand, removes this restriction by allowing arbitrary target sound classes\cite{kavalerov2019universal}. The initial models that aimed to perform this task estimated multiple masks for the input mixture to extract the different sound sources individually \cite{kavalerov2019universal}. More recently, studies such as \cite{liu22w_interspeech, kilgour22_interspeech} proposed models that perform single-target source separation using natural language as the control signal. These models adopted an encoder-decoder architecture conditioned through \textit{Feature-wise Linear Modulation} (FiLM) \cite{perez2018film}. This enabled the source separation models to isolate any arbitrary sound class given an appropriate control signal. While the early LASS models relied on generic sentence encoders, the adoption of CLAP in later work provided semantically aligned language-audio embeddings as the control signal that substantially improved source separation performance \cite{ma2024clapsep, mahmud2024weakly, chung2024language, ma2025language, feng2025hybrid, yang2026clapmamba, liu2024separate, saijo2025leveraging, zhang2026mars}.

A limited number of studies have investigated improving LASS performance for specific subsets of sound classes. The work done in \cite{limited_arbitrary_chen2024mdx} focuses on improving the separated signals of specific target classes after inference, using a generative adversarial network (GAN) to refine the output of the source separation model. Another line of work combines domain partitioning and parameter efficient finetuning (PEFT) to adapt LASS models to multiple subsets of sound classes and subsequently combine the adapted models into a single model \cite{zhang2026domain}. Although these studies consider specific subsets of the sound classes, they do not specifically focus on refining the training data or optimizing the control signal. In this work, we aim to improve LASS performance for a set of defined sound classes through introducing a multi-stage automated data engine for training data curation and an optimized set of control signals for better separation. 

\section{Method}
We propose a framework for adapting LASS models to subsets of sound classes: a multi-stage data engine for constructing and filtering training data, and a control-signal optimization process for refining class-specific CLAP embeddings. Together, these components are used to fine-tune a pretrained LASS model to improve separation performance for target sound classes. An overview of the proposed framework is shown in Fig \ref{fig:overview}.

\subsection{Data Generation and Filtering Pipeline}
\label{subsection:data_engine}
Large-scale audio datasets often contain mislabeled samples or samples where multiple sound classes are present but only one sound class is specified by the label. During the training, the presence of other non-target audio classes in the ground-truth target signal can negatively affect the separation performance of the model. The proposed data engine first uses an audio captioning model which identifies different sound sources in audio samples and provides a detailed caption describing each sample. The caption is then given to an LLM-based judge which is responsible for checking two criteria: (a) Does the audio include only a single sound class? (b) Does the caption fall under any of the defined target sound classes? 

The use of audio captioning models allows the data engine to leverage datasets that do not provide detailed descriptions of their audio samples. This allows us to draw training samples from datasets collected with a wide range of purposes rather than the only the datasets with detailed annotations. For this study, we collected samples from multiple datasets, including AudioSet \cite{gemmeke2017audio}, AudioCaps \cite{audiocaps}, VGG-Sound \cite{chen2020vggsound}, Freesound \cite{fonseca2017freesound}, and BBC Sound Effects platform \cite{bbc_sound_effects}. When existing captions are available, or when multiple captioning models are used, all the generated and pre-existing captions are combined creating a comprehensive description of each audio sample \footnote{The details and the prompts used in this experiment are published in \url{https://amhajavi.github.io/AudioSourceSeparation}}. 

The comprehensive description is then incorporated into the body of the prompt provided to the LLM-based judge, together with the specified subset of target classes. The judge then evaluates whether: (1) The description indicates the presence of only a single sound class or mentions any other classes. (2) Whether that class belongs to the target set.  Samples that do not adhere to these criteria are removed from the candidate pool \footnotemark[\value{footnote}].

\subsection{Control Signal Optimization}

CLAP is trained using a contrastive objective to learn semantically aligned representations of audio and text. As a result, CLAP embeddings are commonly used as a control signal for LASS models, providing an effective means of specifying target sound classes through natural language. However, when adapting a LASS model to the defined sound classes of interest, CLAP embeddings may not adequately capture the semantic coverage of the subcategories within each target class. This can make it difficult for the model to associate certain classes as subcategories of a target class. The proposed control signal optimization provides a solution to this problem through a two-stage process that refines the class-specific CLAP embeddings to improve their ability to represent the target class and its subcategories while maintaining separability from other classes. 

\paragraph*{The first stage} consists of an iterative selection process that selects the largest subset of text prompts subject to a constraint on the cosine distance of the mean text and audio CLAP embeddings. We start with a large set of possible textual descriptions for each sound class and a fixed set of audio samples best representing each sound class. Through this process, we maximize the number of retained prompts per target sound class while maintaining audio-text semantic alignment between the remaining text prompts and the set of audio samples. For a given target sound class, we define $\mathcal{D}^c_T \in \mathbb{R}^{n\times d}$ and $\mathcal{D}^c_A \in \mathbb{R}^{m\times d}$, where $n$ is the number of textual descriptions designated for class $c$, $m$ is a fixed number of audio samples chosen representing the sound class, and $d$ is the size of the CLAP embedding vectors. The subset of the candidate textual prompts for each class $S^c$ is selected through: 
\begingroup
\setlength{\abovedisplayskip}{5pt}
\setlength{\belowdisplayskip}{5pt}
\begin{equation}
        S^c = \underset{S^c \subseteq \{1, \dots, n\}}{\arg\max} \ |S^c| \quad \text{s.t.} \quad \sum_{i=1}^{m}\left(1 - \tfrac {\overline{D^c_T[S]} \cdot D^c_{A_i}}{\|\overline{D^c_T[S]}\| \|D^c_{A_i}\|} \right) \le t
\end{equation}
\endgroup
where $\overline{D^c_T[S]}$ denotes the mean CLAP embeddings of the text prompts indexed by $S$ and $t$ is the threshold of the acceptable cosine distance between the mean text and audio embeddings and is selected by examining the trade-off between the number of retained text prompts and audio-text alignment. After selecting $S^*$, the retained text prompts are used to construct the class-specific control embedding in the second stage.

\paragraph*{The second stage} consists of an optimization procedure to construct the control signals from the selected subset of CLAP embedding candidates. Each class is then represented by an embedding vector, $e_c$, which is the mean vector over all the embedding vectors of $S^c$. The embeddings are then optimized using a stochastic gradient descent (SGD) optimizer with respect to three objectives:
\begin{enumerate}
    \item Orthogonal Normalization ($\mathcal{L}_{\mathrm{Ortho}}$) which aims to maximize the angular distance of the embedding vectors through
    {\footnotesize
        \begin{equation}
            \mathcal{L}_{\mathrm{Ortho}} = \|e_ce_c^T- \mathbb{I}_{n\times n}\|^2
        \end{equation}
    }
    where $\mathbb{I}_{n\times n}$ is the identity matrix. 

    \item Regularization through Deviation ($\mathcal{L}_{\mathrm{Dev}}$) which aims to limit the deviation of the optimized embeddings from the original embeddings $\hat{e}_c$ through 
    
    {\footnotesize
        \begin{equation}
            \mathcal{L}_{\mathrm{Dev}} = \|e_c- \Hat{e}_c\|
        \end{equation}
    }
    
    \item Information Noise-Contrastive Estimation ($\mathcal{L}_{\mathrm{INCE}}$)\cite{zimmermann2024infonce} which performs a contrastive optimization between the embedding and the embeddings belonging to different classes. 
    \begingroup
        \setlength{\abovedisplayskip}{0pt}
        \setlength{\belowdisplayskip}{0pt}
        {\footnotesize
            \begin{equation*}
            s(x,y)=\frac{x^\top y}{\tau}
            \end{equation*}
            \begin{equation*}
            \mathcal{L}_{\mathrm{INCE}}^c= \qquad\qquad\qquad\qquad\qquad\qquad\qquad\qquad\qquad\qquad\quad
            \end{equation*}
            \begin{equation}
                \begin{aligned}
                \mathbb{E}_{e_c,D_A^+,\{D_{A_i}^-\}}
                \left[
                -\ln
                \frac{e^{s(e_c,D_A^+)}}
                {e^{s(e_c,D_A^+)}+\sum_{i=1}^{M}e^{s(e_c,D_{A_i}^-)}}
                \right]
                \end{aligned}
            \end{equation}
        }
    \endgroup
    where $\tau$ is the temperature value which is adopted from \cite{zimmermann2024infonce} and $D_A^+$ is the set of audio clap embeddings belonging to class $c$ and $D_{A_i}^-$ is the set of audio clap embeddings belonging to sound classes other than $c$.
\end{enumerate}

The final objective, $\mathcal{L}_{Total}$, is obtained by a weighted sum of all the three objective functions
\begingroup
    \setlength{\abovedisplayskip}{0pt}
    \setlength{\belowdisplayskip}{0pt}
    \begin{equation}
        \mathcal{L}_\mathrm{Total} = \lambda \mathcal{L}_\mathrm{Ortho} + \omega \mathcal{L}_\mathrm{Dev} + \gamma \sum_{c} \mathcal{L}_{\mathrm{INCE}}^c
    \end{equation}
    \begin{equation*}
        s.t. \qquad \lambda + \omega + \gamma = 1
    \end{equation*}
    where $\lambda$, $\omega$, and $\gamma$ are investigated through experiments. 
\endgroup

\begin{table*}[t]
    \centering
    \label{tab:results}
    \resizebox{0.95\textwidth}{!}{%
    \begin{tabular}{l|ccc|ccc|ccc|ccc}
        \toprule
        & \multicolumn{3}{c|}{SDRi $\uparrow$}
        & \multicolumn{3}{c|}{SI-SDRi $\uparrow$}
        & \multicolumn{3}{c|}{CLAPScore $\uparrow$}
        & \multicolumn{3}{c}{CLAPScore$_A$ $\uparrow$} \\
        \cmidrule(lr){2-4}
        \cmidrule(lr){5-7}
        \cmidrule(lr){8-10}
        \cmidrule(lr){11-13}

        Target class
        & Baseline & Refined & Optimized
        & Baseline & Refined & Optimized
        & Baseline & Refined & Optimized
        & Baseline & Refined & Optimized \\
        \midrule

        Ambient
        & -1.623 & 4.116 & \textbf{4.301}
        & -0.004 & 10.349 & \textbf{10.677}
        & \textbf{0.043} & 0.028 & 0.034
        & 0.559 & 0.708 & \textbf{0.721} \\

        Animal
        & 6.502 & 7.749 & \textbf{7.911}
        & 12.148 & 13.730 & \textbf{13.863}
        & 0.210 & \textbf{0.218} & 0.215
        & 0.763 & 0.744 & \textbf{0.767} \\

        Crowd
        & 2.507 & \textbf{4.060} & 4.026
        & 8.449 & \textbf{10.596} & 10.561
        & \textbf{0.145} & 0.120 & 0.123
        & 0.794 & 0.830 & \textbf{0.843} \\

        Music
        & 5.378 & 6.142 & \textbf{6.232}
        & 11.475 & 12.316 & \textbf{12.435}
        & \textbf{0.207} & 0.197 & 0.198
        & \textbf{0.780} & 0.753 & 0.765 \\

        Nature
        & 0.071 & 4.285 & 4.083
        & 5.050 & \textbf{10.547} & 10.366
        & \textbf{0.146} & 0.142 & 0.144
        & 0.766 & 0.866 & \textbf{0.869} \\

        Speech
        & 6.473 & 7.762 & \textbf{7.844}
        & 12.418 & 14.223 & \textbf{14.308}
        & 0.205 & \textbf{0.226} & 0.222
        & 0.711 & \textbf{0.719} & 0.694 \\

        Wind 
        & 5.410 & 6.292 & \textbf{6.293}
        & 11.289 & \textbf{12.368} & 12.368
        & 0.219 & 0.220 & \textbf{0.225}
        & 0.742 & \textbf{0.805} & 0.798 \\

        \midrule

        \textbf{Average}
        & 2.753 & 5.545 & \textbf{5.610}
        & 7.386 & 11.798 & \textbf{11.904}
        & \textbf{0.148} & 0.143 & 0.145
        & 0.707 & 0.763 & \textbf{0.770} \\

        \bottomrule
    \end{tabular}%
    }
    \caption{\footnotesize Performance comparison of the baseline, refined, and optimized control signals across target sound classes.
    Best results for each target class and metric are shown in bold.
    $\uparrow$ indicates that higher values are better.}
\end{table*}

\section{Experiments and Results}

We select 7 classes of "\emph{Ambient}", "\emph{Animal}", "\emph{Crowd}", "\emph{Music}", "\emph{Nature}", "\emph{Speech}", and "\emph{Wind}" to evaluate the improvement in source separation performance achieved by our proposed framework. The selection of these classes was based on the AudioSet Ontology\cite{gemmeke2017audio}, the frequency of appearance in real-world videos, and inspired by the current commercially available options with similar functionality (e.g. Google Audio Magic Eraser and Samsung Audio Eraser). We applied the automated data engine on each dataset independently, as described in Section \ref{subsection:data_engine}, with a target of 20K samples per sound class. The data engine generated more than 20K samples for the \emph{speech} and \emph{music} classes. In order to maintain a balanced class distribution, we randomly selected 20K samples from each class. In contrast, for classes such as \emph{Crowd} and \emph{Wind}, the combined number of audio samples across all the datasets did not reach 20K. We therefore supplemented the remaining samples using audio generative models and handcrafted techniques \footnotemark[\value{footnote}]. 

Since existing audio source separation models use a wide range of datasets for evaluation and there are no standard test set used consistently across the models, we selected a simple baseline architecture for our experiments. Specifically, we use AudioSep \cite{liu2024separate} model that introduces minimal architectural components beyond CLAP embeddings and FiLM layers as the control signal mechanism for single-channel source extraction. This choice allows us to evaluate the improvement in source separation through the proposed control signal optimization without additional architectural factor that could impact the comparison. We use Signal-to-Distortion Ratio improvement (SDRi) and the scale invariant version of it, SI-SDRi, as signal-level objective metrics to evaluate the performance of the source separation. We also use CLAPScore and CLAPScore$_A$ as metrics to measure the semantic alignment of the extracted audio with respect to the caption (The sound of "Sound Class") and the audio samples of the same sound class in test set. We extend our evaluation to a subjective preference test among 17 participants to compare the sound separation quality of the models trained using our refined dataset and optimized control signal against the baseline model and the commercial models from Google and Samsung. We opted to do a listening test whereas the objective tests against the commercial options were not possible. This was due to the heavy processing applied on the audio signal when passing through the mobile apps which results in SDRi and SI-SDRi values of -7.668 and -34.914. 

\begin{figure}[!t]
    \centering
    \includegraphics[width=0.9\linewidth]{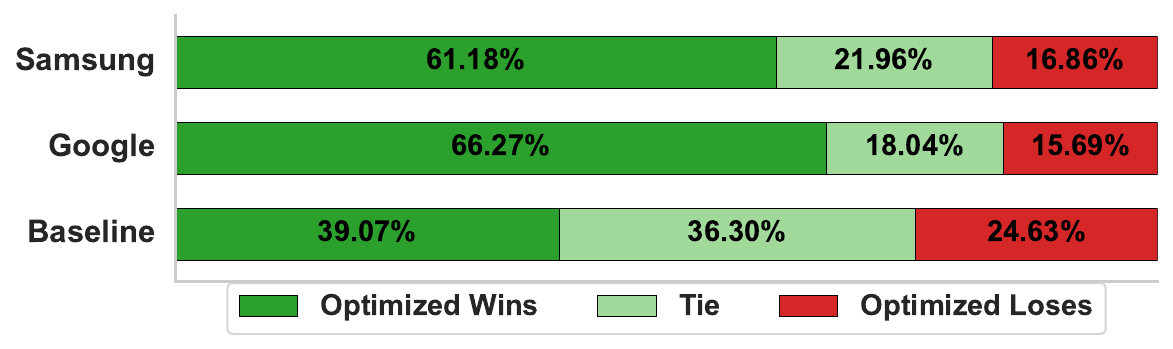}
    \caption{\footnotesize Subjective preference evaluation of the separation quality for the model trained using optimized control signal.}
    \label{fig:subjective-evaluation}
\end{figure}

As shown in Table~\ref{tab:results}, both SDRi and SI-SDRi consistently improve from the baseline model to the refined model, with further improvement achieved by the optimized model. The improvements from the baseline to the refined model indicate the benefits of the automated data engine for filtering and cleaning the training data, while further improvements from the refined to the optimized model highlight the effect of using the control signals obtained by the proposed optimization process. In contrast, CLAPScore shows minimal variation across three models. This may be attributed to the limited initial alignment between the CLAP embeddings and the target sound classes which further highlights the need for an optimization of CLAP-based control signals for source separation in scenarios with a limited set of target sound classes. CLAPScore$_A$, however, shows improvement from the baseline through the refined and optimized models, indicating the improved semantic similarity between the extracted audio and the target classes. 

Figure~\ref{fig:subjective-evaluation} presents the results of the subjective evaluation, comparing the optimized model against the baseline model and the commercial options from Google and Samsung. Compared to the baseline model, the optimized model is preferred in 39.07\% of the trials, has the loss rate of 24.63\%, and 36.30\% of the comparisons result in a tie. Our comparison with the commercial options also shows an overwhelming preference towards our model compared to both the models from Google and Samsung. Overall, the optimized model is preferred more frequently than it is disfavored in the subjective evaluations with both the baseline model and the commercial options. These results suggest that the proposed control-signal optimization provides a perceptually meaningful improvement in the separation quality.

\section{Conclusion}
Adapting LASS models to a defined set of general classes faces challenges such as training data procurement and lack of an effective control signal for source separation. We propose a framework consisting of an automated data engine and an optimization process for CLAP-based control signals to overcome these challenges. Through both subjective and objective tests we show that our framework is effective in adapting pretrained LASS models which outperforms both the baseline and commercial options. As future directions, we aim to explore other optimization methods for CLAP-based control signals and further refine the text prompt selection process.

\bibliographystyle{IEEEbib}

\footnotesize
\bibliography{strings,refs}

\end{document}